# Are Yao Graph and Theta Graph Void Free?

Weisheng Si
School of Computing, Engineering, and Mathematics
University of Western Sydney
Sydney, Australia
w.si@uws.edu.au

*Abstract*—Greedy Forwarding algorithm is a widely-used routing algorithm for wireless networks. However, it can fail if the wireless network topologies contain void scenarios (or simply called voids). Since Yao Graph and Theta Graph are two types of geometric graphs exploited to construct wireless network topologies, this paper studies whether these two types of graphs can contain voids. Specifically, this paper shows that when the number of cones in Yao Graphs or Theta Graphs is less than 6, Yao Graphs and Theta Graphs can have voids, but when the number of cones equals or exceeds 6, Yao Graphs and Theta Graphs are free of voids.

*Index Terms*—greedy forwarding, Yao graph, Theta graph, void scenario, wireless networks.

## I. Introduction

With the use of GPS or other localization techniques, the position information of wireless nodes becomes available in many wireless sensor networks, wireless mesh networks, or vehicular ad hoc networks. For these networks, the network topology can be modeled by a geometric graph $G(V, E)$ in which each node in $V$ is associated with $(x, y)$-coordinates, and each edge in $E$ represents a connection between two nodes and has a weight equal to the Euclidean distance between these two nodes.

An important routing algorithm under the above geometric graph model is the Greedy Forwarding algorithm [1]: when a node $u$ forwards a packet with destination node $t$, $u$ sends this packet to its neighbor that has the smallest Euclidean distance to $t$. Here, two nodes $u$ and $v$ are said to be each other's neighbor if the edge $uv$ is present in the graph. The advantages of the Greedy Forwarding algorithm include: (1) low computation overhead at a node, (2) low space overhead for a packet, and (3) the capability to achieve short paths.

However, Greedy Forwarding does not succeed on a graph that contains the *void* scenario (shortened as *void*) [1], in which for a certain destination $t$, a node does not have a neighbor with a smaller distance to $t$ than its own distance to $t$. Thus, whether a geometric graph contains voids becomes an important property to study. To the convenience of study, we formally define the concept *void-free* as follows. Let $d(a, b)$ denote the Euclidean distance between node $a$ and node $b$; if for any node pair $(u, v)$ in a geometric graph $G$, $u$ always has a neighbor $w$ such that $d(w, v) < d(u, v)$, $G$ is said to be *void-free*.

The void-free property has been studied for several types of geometric graphs used in wireless networks such as Relative Neighborhood Graph [2], Gabriel Graph [3], and Delaunay Triangulation [4]. In [5], Delaunay Triangulations are shown to be *void-free*. In [6], counter-examples are given to show that Relative Neighborhood Graphs and Gabriel graphs are not void-free for certain node sets. However, for Yao Graphs [7] and Θ-Graphs (or Theta Graphs) [8], which are also leveraged in many works [9-12] to construct network topologies, no results exist on the void-free property yet. Therefore, this paper investigates on this aspect. For later reference in this paper, the definitions of Yao Graph and Θ-Graph are stated below.

Given a set $V$ of nodes on the plane, the directed Yao Graph with an integer parameter $k$ ($k \geq 1$) on $V$ is obtained as follows. For each node $u \in V$, starting from the direction of positive y-axis, draw $k$ equally-spaced rays $l_1, l_2, \ldots, l_k$ originating from $u$ in clockwise order (see Fig. 1 (a) below). These rays divide the plane into $k$ cones, denoted by $c(u, 1), c(u, 2), \ldots, c(u, k)$ respectively in clockwise order. To avoid overlapping at boundaries, it is required that the area of $c(u, i)$, where $i=1, \ldots, k$, excludes the ray $l_{i \% k}$ but includes the ray $l_{(i+1) \% k}$. In each cone of $u$, draw a directed edge from $u$ to its closest node by Euclidean distance in that cone. Ties are broken arbitrarily. These directed edges will form the edge set of the directed Yao graph on $V$. The undirected Yao Graph (or simply Yao Graph) on $V$ is obtained by ignoring the directions of the edges. Note that if both edge $uv$ and $vu$ are in the directed Yao graph, only one edge $uv$ exists in the Yao graph. Fig. 1 (b) gives an example of Yao graph with $k=5$.

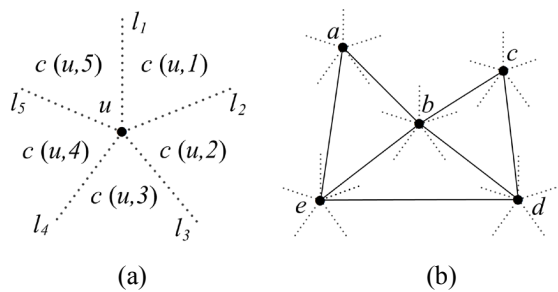

(a)      (b)

Fig. 1. Cones and an example of Yao Graph for $k=5$.

Similar to Yao graph, the undirected Θ-Graph (or simply Θ-Graph) is also obtained by letting each node $u \in V$ select a 'closest' node in each of its cones to have an edge. The only difference is that 'closest' in Θ-Graph means the smallest projection distance onto the bisector of that cone, not the direct Euclidean distance. For convenience, we denote Yao Graph and Θ-Graph with parameter $k$ as $Y_k$ and $Θ_k$ hereafter.

Note that this paper assumes that transmission ranges of wireless nodes are long enough such that complete Yao Graphs or Θ-Graphs can be established. The rationales for this assumption are that: (1) in most cases, the links needed by complete Graphs or Θ-Graphs have lengths within the transmission ranges of wireless nodes; (2) the increasing use of directional antennas on wireless nodes enables long transmission range; (3) by considering complete Yao Graphs or Θ-Graphs, basic results can be obtained and these results can also be applied to other research areas such as transportation networks and robotics.

The main contributions of this paper are as follows:
- When $1 \leq k \leq 5$, $Y_k$ and $Θ_k$ are not void-free for certain node sets on the plane.
- When $k \geq 6$, $Y_k$ and $Θ_k$ are void-free for any node set on the plane.

## II. COUNTER EXAMPLES WHEN $1 \leq K \leq 5$

When $1 \leq k \leq 5$, we give counter examples to show that $Y_k$ and $Θ_k$ are not always void-free. Specifically, we establish the following two propositions.

**Proposition 1**. When $1 \leq k \leq 5$, some node sets exist such that $Y_k$ are not void-free.

Proof: When $1 \leq k \leq 3$, we give a counter-example node set $V_0$ with four nodes $u$, $v$, $a$, and $b$ as shown in Fig. 2, where the dotted auxiliary circle $C_v$ is centered at $v$ and has radius $d(u, v)$, and the dotted auxiliary lines emanating from each node are only drawn for $k = 3$.

When $k = 1$, $Y_1$ on $V_0$ is actually the nearest neighbor graph on $V_0$ [13], thus only having two edges $ua$ and $vb$. As node $a$ lies outside the circle $C_v$, $u$ does not have a neighbor with a shorter distance to $v$. So $Y_1$ is not void-free on $V_0$.

When $k = 2$ and 3, $Y_2$ and $Y_3$ on $V_0$ are the same and depicted by solid lines in Fig. 2. Again, node $u$ does not have a neighbor with a shorter distance to $v$. So $Y_2$ and $Y_3$ are not void-free on $V_0$.

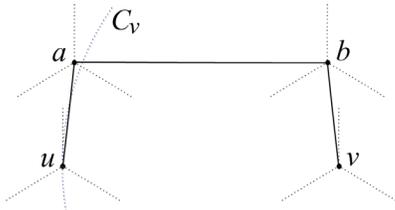

Fig. 2. The counter-example node set $V_0$ on which Yao Graphs are not void-free for $k = 1, 2, 3$.

When $k = 4$, we give a counter-example node set $V_1$ with six nodes $u$, $v$, $a$, $b$, $c$, and $d$ as shown in Fig. 3. In this figure, the dotted circle $C_v$ is centered at $v$ and has radius $d(u, v)$. The resulting $Y_4$ on $V_1$ is depicted by solid lines. Since nodes $a$ and $b$ are outside the circle $C_v$, node $u$ does not have a neighbor with shorter distance to $v$. So both $Y_4$ is not void-free on $V_1$.

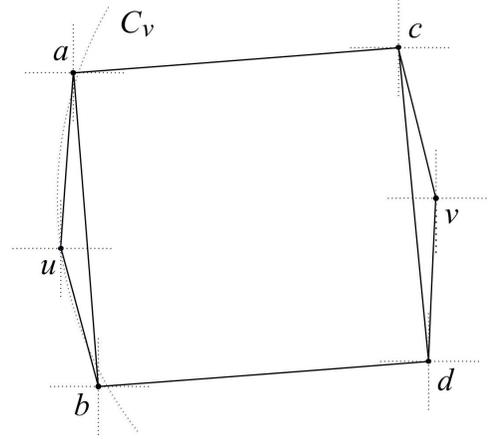

Fig. 3. The counter-example node set $V_1$ on which Yao Graphs are not void-free for $k = 4$.

When $k = 5$, we give a counter-example node set $V_2$ with six nodes $u$, $v$, $a$, $b$, $c$, and $d$ as shown in Fig. 4. Their positions have the following relationship: $v$ is on the ray $l_2$ originating from $u$, so $v$ is located inside $c(u, 1)$; $d$ is located inside $c(v, 4)$; $b$ is located inside $c(u, 3)$, $c(d, 4)$, and $c(v, 4)$; $c$ is located inside $c(u, 2)$, $c(d, 3)$ and $c(v, 3)$; $a$, $b$, and $c$ are outside the circle $C_v$ which is centered at $v$ and has radius $d(u, v)$. With this node placement, the resulting $Y_5$ on $V_2$ is shown by solid lines in Fig. 4. As nodes $a$, $b$, and $c$ are outside the circle $C_v$, node $u$ does not have a neighbor with a shorter distance to $v$. So $Y_5$ is not void-free on $V_2$. □

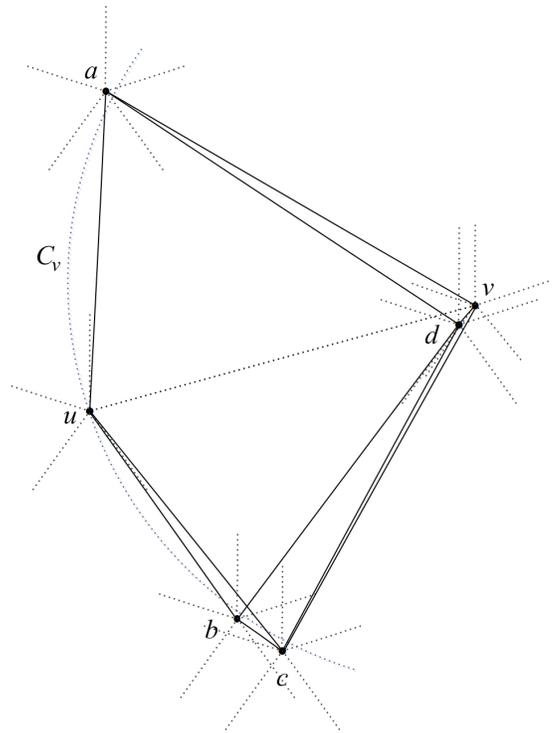

Fig. 4. The counter-example node set $V_2$ on which Yao Graphs are not void-free for $k = 5$.

It is not hard to verify that the above counter examples $V_0$, $V_1$, and $V_2$ also apply to $\Theta$-Graph. That is, for $k = 1, 2, 3$, $\Theta_k$ on $V_0$ are the same as $Y_k$ on $V_0$; and for $k = 4$, $\Theta_4$ on $V_1$ is the same as $Y_4$ on $V_1$; and for $k = 5$, $\Theta_5$ on $V_2$ is the same as $Y_5$ on $V_2$. Therefore, we can have the following proposition.

**Proposition 2**. When $1 \leq k \leq 5$, some node sets exist such that $\Theta_k$ are not void-free.

Moreover, we verified our counter examples by experiments. Specifically, by using the online tool 'Visualization of Spanners' [14], we confirmed that our counter examples $V_0$, $V_1$, and $V_2$ indeed give all the $Y_k$'s and $\Theta_k$'s as depicted in the previous figures.

### III. PROOFS FOR VOID-FREE WHEN $K \geq 6$

When $k \geq 6$, we show that $Y_k$ and $\Theta_k$ are void-free for any node set by proving the following two propositions.

**Proposition 3**. When $k \geq 6$, $Y_k$ are void-free for any node set.

Proof: This proof is done by providing the following method through which, for any two given nodes $u$, $v$ in a $Y_k$ with $k \geq 6$, $u$ can find a neighbor $w$ in $Y_k$ such that $d(w, v) < d(u, v)$. To start, node $v$ must reside in one cone of $u$ (see Fig. 5). According to the definition of $Y_k$, $u$ connects to one of its closest neighbors in that cone. Denote this neighbor by $w$, we next show that $d(w, v) < d(u, v)$. If $u$, $w$, and $v$ are collinear, this inequality obviously holds. Otherwise, we can draw a triangle connecting nodes $u$, $w$, and $v$. When $k \geq 6$, the angle of a cone is no more than $\pi/3$. Because $w$ and $v$ cannot fall on different boundaries of a cone at the same time (due to cone's definition), we have $\angle wuv < \pi/3$ and also $\angle uvw + \angle uwv > 2\pi/3$. Since $d(u, w) \leq d(u, v)$, we have $\angle uvw \leq \angle uwv$. Since we already know $\angle uvw + \angle uwv > 2\pi/3$, we have $\angle uwv > \pi/3$. Thus, $\angle uwv > \angle wuv$. Since edge $uv$ opposites $\angle uwv$ and edge $wv$ opposites $\angle wuv$, we have $d(u, v) > d(w, v)$. □

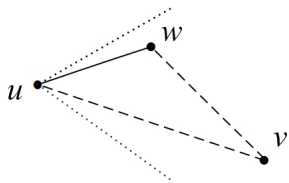

Fig. 5. Proof for Yao Graph for $k \geq 6$.

**Proposition 4**. When $k \geq 6$, $\Theta_k$ are void-free for any node set.

Proof: This proof is also done by providing a method by which for any given $u$, $v$ in a $\Theta_k$ with $k \geq 6$, $u$ can always find a neighbor $w$ in $\Theta_k$ such that $d(w, v) < d(u, v)$. Similar to the above proof, node $v$ must reside in one cone of $u$ (see Fig. 6). According to the definition of $\Theta_k$, $u$ connects to its neighbor that has the shortest projection distance on the bisector of that cone. Denote this neighbor by $w$, and the projection points of $w$ and $v$ on the bisector by $w'$ and $v'$ respectively. If $u$, $w$, and $v$ are collinear, obviously we have $d(w, v) < d(u, v)$. Otherwise, nodes $u$, $w$, and $v$ form a triangle. There can be two cases as illustrated in Fig. 6: in Case 1, $\angle uwv$ faces the bisector; in Case 2, $\angle uwv$ does not face the bisector.

For Case 1, when $k \geq 6$, the angle between the boundary and the bisector is no more than $\pi/6$. Thus we have $\angle uww' \geq \pi/2 - \pi/6 = \pi/3$. Since $d(u, w') \leq d(u, v')$, we have $\angle uww' \leq \angle uwv$. Because $w$ and $v$ cannot fall on different boundaries of a cone (due to cone's definition), we have $\angle wuv < \pi/3$. Thus, $\angle wuv < \angle uwv$. Since edge $wv$ opposites $\angle wuv$ and edge $uv$ opposites $\angle uwv$, we have $d(w, v) < d(u, v)$.

For Case 2, it is easy to show $d(u, w) < d(u, v)$. Then, following the proof for Proposition 3, we can obtain $d(w, v) < d(u, v)$. □

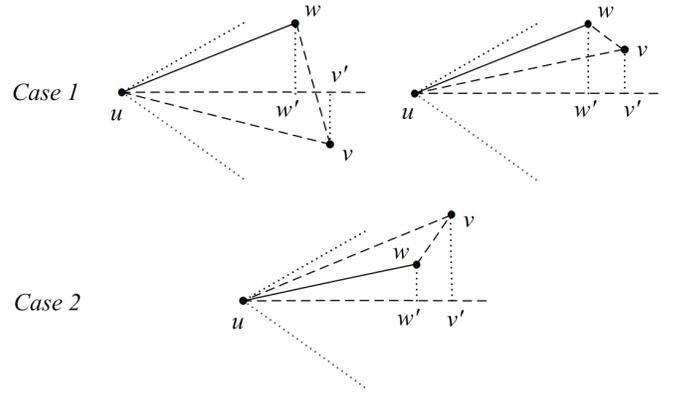

Fig. 6. Proof for Theta Graph for $k \geq 6$.

Finally, we note that the above two proofs still work when a node encounters ties to select 'closest' neighbors in a cone to construct an edge in Yao Graph or $\Theta$-Graph.

### IV. CONCLUSIONS

Desired by the Greedy Forwarding algorithm, the void-free property becomes an important issue to investigate for wireless networks. This paper studies whether Yao Graph and $\Theta$-Graph are void-free for different $k$, where $k$ represents the number of cones. Specifically, this paper shows that (1) when $1 \leq k \leq 5$, $Y_k$ and $\Theta_k$ may not be void-free, and the counter examples given are verified by experiments; (2) when $k \geq 6$, $Y_k$ and $\Theta_k$ are void-free for any node set.